\documentclass[a4paper,15pt]{article}
\usepackage{graphicx}
\usepackage{float} 
\usepackage[T1]{fontenc}
\usepackage{authblk}
\usepackage{amsmath}
\usepackage{caption}
\usepackage{geometry}
\date{}
\title{Huygens' Principle Reveals Dispersion in Inhomogeneous Media \vspace{+5em}}
\author[ ]{Li Mingcong}
\author[ ]{ Zhao Zhenming\thanks{Corresponding author:zhaozhenming@cust.edu.cn}}
\affil[ ]{\slshape \small School of Physics}
\affil[ ]{\slshape \small Changchun University of Science and Technology}
\affil[ ]{\slshape \small Changchun, 130022, China \vspace{+5em}}
\begin{document}
\captionsetup[figure]{labelfont={bf},labelformat={default},labelsep=period,name={Fig.}}
\maketitle
\let\sss= \scriptscriptstyle

\begin{abstract}
Dispersion is an important factor of optical materials. Due to the effect of techniques and equipment in the manufacturing process of optical materials, the inhomogeneity of the material may be caused. In this paper, microsphere optical media are used to replace the inhomogeneous zones, and Huygens’ principle is used to study the dispersion caused by the material inhomogeneity. First, we study the effect of a single inhomogeneous zone, and then the effect of a thin medium with a large number of inhomogeneous zones. It is deduced that the dispersion law of a macro-optical medium is also consistent with Cauchy formula. Finally, it is pointed out that Huygens' principle is suitable for studying the interaction between light and particles.

\end{abstract}
\thispagestyle{empty}
\newpage

\section{Introduction}
Dispersion is a common optical phenomenon. As early as 1666, Newton studied dispersion using prisms\cite{newton1952opticks} In 1837, Cauchy gave his dispersion formula\cite{cauchy1836memoire}. In 1872, Sellmeier developed Cauchy dispersion formula and put forward the Sellmeier equation, which is also an empirical formula\cite{sellmeier1872ueber}. Dispersion belongs to light-matter interaction. According to current theory, the dispersive properties of an optical medium depend on the interaction between the incident light and the electrons in the medium, which is a typical interaction between light and microscopic particles. From Huygens' principle, where there are waves, there are wavefronts and wavelets\cite{huygens1912treatise}. With the understanding about the essence of Huygens' principle\cite{zhang2018study}, it was found that Huygens' principle is feasible for studying the interaction between light and microscopic particles, and has been applied to the research of Rayleigh scattering\cite{zhang2020huygens}, atomic stimulated radiation\cite{zhao2016phase} and photon absorption\cite{zhao2018huygens}. In this work, we use Huygens’ principle to analyze the microscopic interaction between light and a medium and the propagation of light in inhomogeneous media. It is found that the inhomogeneity of optical media can lead to dispersion.

\section{The influence of a single inhomogeneous zone on light propagation}
Optical uniformity is an important index for optical materials. In common optical media, there are local inhomogeneity in density or composition, which leads to fluctuations in refractive index of materials. It is important to study the effect of refractive index fluctuations on light propagation. We use a microsphere with refractive index n and radius R to describe the local refractive index fluctuations in a single inhomogeneous zone.	
\begin{figure}[H] 
\centering 
\includegraphics[width=0.7\textwidth]{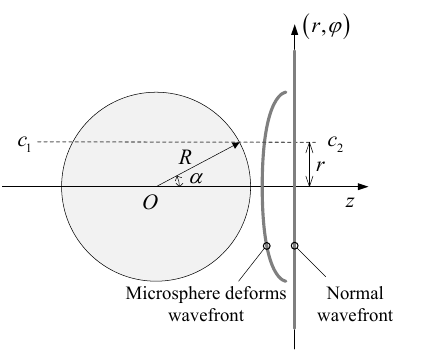} 
\caption{Influence of a microsphere on the wavefront} 
\label{Fig.1} 
\end{figure}	
In Fig.\ref{Fig.1}, the wavefront of the plane wave propagating along the $z$ axis is changed by the microsphere, and the deformation degree of the wavefront is related to the wavelength. When the light travels along the $c_1c_2$ path, the influence of the microsphere on the optical path and phase is,
\begin{equation}\label{eq.1}
\delta \left (r,\varphi   \right ) =k\left ( n-1 \right ) \cdot 2R\cos\alpha 
\end{equation}
where $k$ is the incident light wave number. The wavelength of the incident light is $\lambda$ and the amplitude is $A$, so complex amplitude distribution in the plane is given by
\begin{equation}\label{eq.2}
	U\left ( r,\varphi ,0 \right ) =At\left (r,\varphi\right  )
\end{equation}
where $t(r,\varphi)$ is the transmittance function, which is
\begin{equation}\label{eq.3}
	t(r,\delta )=\left\{\begin{matrix}
		1 &r>R\\
		e^{i\delta} &r\le R
	\end{matrix}\right.
\end{equation}
\begin{figure}[H] 
	\centering 
	\includegraphics[width=0.7\textwidth]{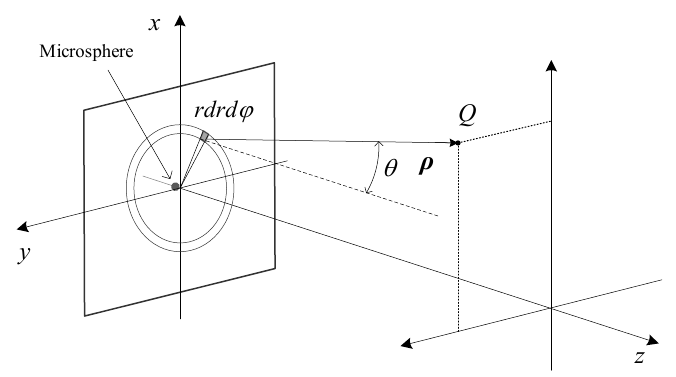} 
	\caption{The $r\mathrm{d}r\mathrm{d}\varphi$ area element emits Huygens wavelet} 
	\label{Fig.2} 
\end{figure}
In Fig.\ref{Fig.2}, according to the parameters of the incident light and the microsphere, the complex amplitude of the light at $z = 0$ can be obtained. Using Huygens' principle, the amplitude of the light at point Q is
\begin{equation}\label{eq.4}
	 U_{Q} =\frac{1}{i\lambda }\int\limits_{0}^{\infty } \int\limits_{0}^{2\pi }U(r,\varphi ,0)\frac{e^{ik\rho } }{\rho}K(\theta )r\mathrm{d}r\mathrm{d}\varphi  
	 =\frac{A}{i\lambda }\int\limits_{0}^{\infty } \int\limits_{0}^{2\pi }t(r,\varphi )\frac{e^{ik\rho } }{\rho}K(\theta )r\mathrm{d}r\mathrm{d}\varphi  
\end{equation}
where, $r\mathrm{d}r\mathrm{d}\varphi$ is the area element, $\rho$ is the distance between the area element and point $Q$, and $K(\theta )$ is the inclination factor. Expanding the transmittance function:
\begin{equation}\label{eq.5}
t(r,\varphi )=\left\{\begin{matrix}
	1 &r>R \\
	e^{i\rho }=1+i\delta -\frac{\delta^2}{2}-\frac{i\delta^3}{6}+\frac{\delta^4}{24}+\frac{i\delta^5}{120}+\varepsilon (\delta )  &r\le R
\end{matrix}\right.	
\end{equation}
Substituting Eq.(\ref{eq.5}) into Eq.(\ref{eq.4}) and retaining the first six terms gives:
\begin{equation}\label{eq.6}
	U_{Q}\approx Ae^{ikz}+\frac{A}{i\lambda}\int\limits_{0}^{R}\int\limits_{0}^{2\pi}[i\delta-\frac{\delta ^{2}}{2}-\frac{i\delta ^{3}}{6}+\frac{\delta ^{4}}{24}+\frac{i\delta ^{5}}{120}] \frac{e^{ik\rho } }{\rho}K(\theta )r\mathrm{d}r\mathrm{d}\varphi  
\end{equation}
The range of the wave front affected by the particle is very small. Therefore $K(\theta)e^{ik\rho}/\rho$ can be regarded as a constant, and Eq.\ref{eq.6} can be written as:
\begin{equation}\label{eq.7}
	\begin{aligned}
	U_{Q}&=Ae^{ikz}+\frac{A}{i\lambda}K(\theta ) \frac{e^{ik\rho } }{\rho}\int\limits_{0}^{R}\int\limits_{0}^{2\pi}[i\delta-\frac{\delta ^{2}}{2}-\frac{i\delta ^{3}}{6}+\frac{\delta ^{4}}{24}+\frac{i\delta ^{5}}{120}]r\mathrm{d}r\mathrm{d}\varphi  \\
	&=U_{0}+\frac{Ake^{ik\rho}}{i\rho}K(\theta )\int\limits_{0}^{R}[i\delta-\frac{\delta ^{2}}{2}-\frac{i\delta ^{3}}{6}+\frac{\delta ^{4}}{24}+\frac{i\delta ^{5}}{120}]r\mathrm{d}r\\
	\end{aligned}
\end{equation}
In Eq.(\ref{eq.7}), the first term $Ae^{ikz}$ is the complex amplitude of the incident light at point $Q$, and the following five terms are the complex amplitudes of lights diffracted by the microsphere. We denote these five terms as $U_{1}~U_{5}$ in turn. 
\begin{equation}\label{eq.8}
	U_{1}=\frac{Ake^{ik\rho}}{i\rho}K(\theta )\int\limits_{0}^{R}[i\delta]r\mathrm{d}r
\end{equation}
\begin{equation}\label{eq.9}
	U_{2}=\frac{Ake^{ik\rho}}{i\rho}K(\theta )\int\limits_{0}^{R}[-\frac{\delta ^{2}}{2}]r\mathrm{d}r
\end{equation}
\begin{equation}\label{eq.10}
	U_{3}=\frac{Ake^{ik\rho}}{i\rho}K(\theta )\int\limits_{0}^{R}[-\frac{i\delta ^{3}}{6}]r\mathrm{d}r
\end{equation}
\begin{equation}\label{eq.11}
	U_{4}=\frac{Ake^{ik\rho}}{i\rho}K(\theta )\int\limits_{0}^{R}[\frac{\delta ^{4}}{24}]r\mathrm{d}r
\end{equation}
\begin{equation}\label{eq.12}
	U_{5}=\frac{Ake^{ik\rho}}{i\rho}K(\theta )\int\limits_{0}^{R}[\frac{i\delta ^{5}}{120}]r\mathrm{d}r
\end{equation}
Using the phase shift expression given in Eq.(\ref{eq.1}) and according to $r=R\sin \alpha$, $\mathrm{d}r=R\cos \alpha\mathrm{d}\alpha$, and then integrating we can get
\begin{equation}\label{eq.13}
	U_{1}=\frac{2(n-1)R^{3}k^{2}}{3}\frac{Ake^{ik\rho}}{i\rho}
\end{equation}
\begin{equation}\label{eq.14}
	U_{2}=\frac{i(n-1)^{2}R^{4}k^{3}}{2}\frac{Ake^{ik\rho}}{i\rho}
\end{equation}
\begin{equation}\label{eq.15}
	U_{3}=-\frac{4(n-1)^{3}R^{5}k^{4}}{15}\frac{Ake^{ik\rho}}{i\rho}
\end{equation}
\begin{equation}\label{eq.16}
	U_{4}=-\frac{i(n-1)^{4}R^{6}k^{5}}{9}\frac{Ake^{ik\rho}}{i\rho}
\end{equation}
\begin{equation}\label{eq.17}
	U_{5}=\frac{4(n-1)^{5}R^{7}k^{6}}{105}\frac{Ake^{ik\rho}}{i\rho}
\end{equation}
The phases of $U_{1}$, $U_{3}$, $U_{5}$ and $U_{0}$ are the same. $i=e^{i\pi/2}$ in $U_{2}$ indicates that the phase of $U_{2}$ is $\pi/2$ behind that of $U_{0}$, and the $-i=e^{-i\pi/2}$ in $U_{4}$ shows that its phase is $\pi/2$ ahead of that of $U_{0}$.

\section{Property of inhomogeneous thin medium}
Usually, there are a lot of non-uniform refractive index zones in an optical medium. They are very tiny in dimension, but large in number, and evenly distributed in a fine optical medium. To study the dispersion of such a medium, we start with a layer of the medium.
\begin{figure}[H] 
	\centering 
	\includegraphics[width=0.7\textwidth]{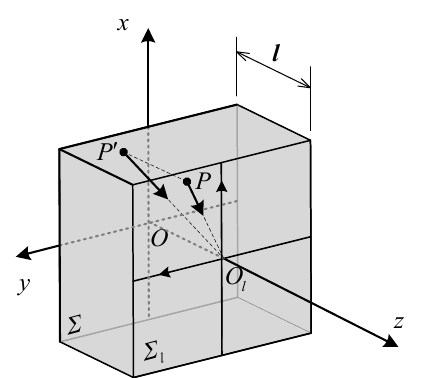} 
	\caption{Microsphere diffraction and equivalent substitution in a thin monolayer medium} 
	\label{Fig.3} 
\end{figure}
In Fig.\ref{Fig.3}, the thickness of the monolayer medium is $l$, and the incident light propagates along the $Z$-axis. The light is diffracted by the microsphere located at point $P$. According to equations (\ref{eq.13}) - (\ref{eq.17}), the amplitude of the diffraction light at point $O_l(0,0,l)$ is:
\begin{equation}\label{eq.18}
	\begin{aligned}
	U_{O_{l}}=&\frac{AK(\theta)e^{ik\rho_{\sss P}}}{i\rho_{\sss P}}[\frac{2k^{2}(n-1)R^{3}}{3}+\frac{i3k^{3}(n-1)^{2}R^{4}}{2} \\
	&-\frac{4k^{4}(n-1)^{3}R^{5}}{15}-\frac{ik^{5}(n-1)^{4}R^{6}}{9}+\frac{4k^{6}(n-1)^{5}R^{7}}{105}]
\end{aligned}
\end{equation}
where $\rho_{\sss P}$ represents the distance of $\overline{PO_{l}} $.\\
When a large number of microspheres exist in the medium, the incident light will be diffracted by each microsphere. Then the total amplitude at point $O_{l}$ is
\begin{equation}\label{eq.19}
	\begin{aligned}
	U_{O_{l}}=&\sum_{\rho_{\sss P}} \frac{AK(\theta)e^{ik\rho_{\sss P}}}{i\rho_{\sss P}}[\frac{2k^{2}(n-1)R^{3}}{3}+\frac{i3k^{3}(n-1)^{2}R^{4}}{2} \\
	&-\frac{4k^{4}(n-1)^{3}R^{5}}{15}-\frac{ik^{5}(n-1)^{4}R^{6}}{9}+\frac{4k^{6}(n-1)^{5}R^{7}}{105}]
    \end{aligned}
\end{equation}
In Eq. (\ref{eq.19}), the particles are thought to be concentrated on the $xy$ plane for simplicity. Point $P(r_{\sss P},\varphi_{ \sss P},z_{\sss  P})$ is equivalent to point  $P^{\prime}(r_{\sss P},\varphi_{ \sss P},0)$, then
\begin{equation}\label{eq.20}
	\begin{aligned}
		U_{O_{l}}=&\sum_{\rho_{\sss P}} \frac{AK(\theta)e^{ik\rho_{\sss P^{\prime}}}}{i\rho_{\sss P^{\prime}}}[\frac{2k^{2}(n-1)R^{3}}{3}+\frac{i3k^{3}(n-1)^{2}R^{4}}{2} \\
		&-\frac{4k^{4}(n-1)^{3}R^{5}}{15}-\frac{ik^{5}(n-1)^{4}R^{6}}{9}+\frac{4k^{6}(n-1)^{5}R^{7}}{105}]
	\end{aligned}
\end{equation}
where, $\rho_{\sss P^{\prime}}$ represents the length of $P^{\prime}O_l$. Let the density of the microspheres be $N$. Then the number of the microspheres on per unit area of the $xy$ plane is $Nl$, and the sum in Eq. (\ref{eq.20}) can be transformed into an integral:
\begin{equation}\label{eq.21}
	\begin{aligned}
		U_{O_{l}}=&[\int\limits_{0}^{\infty }\int\limits_{0}^{2\pi }\frac{NlAK(\theta)e^{ik\rho_{\sss P^{\prime}}}}{i\rho_{\sss P^{\prime}}}r\mathrm{d}r\mathrm{d}\varphi ][\frac{2k^{2}(n-1)R^{3}}{3}+\frac{i3k^{3}(n-1)^{2}R^{4}}{2} \\
		&-\frac{4k^{4}(n-1)^{3}R^{5}}{15}-\frac{ik^{5}(n-1)^{4}R^{6}}{9}+\frac{4k^{6}(n-1)^{5}R^{7}}{105}]
	\end{aligned}
\end{equation}
According to Fresnel wave zone method\cite{bohren1985colors}, we can have $\int\limits_{0}^{\infty }\int\limits_{0}^{2\pi }\frac{AK(\theta)e^{ik\rho_{\sss P^{\prime}}}}{i\rho_{\sss P^{\prime}}}r\mathrm{d}r\mathrm{d}\varphi=i\lambda Ae^{ikl}$. Then Eq.(\ref{eq.21}) can be rewritten as
\begin{equation}\label{eq.22}
	\begin{aligned}
		U_{O_{l}}=&\frac{i4\pi k(n-1)R^{3}ANl}{3}-\frac{6\pi k^{2}(n-1)^{2}R^{4}ANl}{2}-\frac{i8\pi k^{3}(n-1)^{3}R^{5}ANl}{15}\\
		&+\frac{2\pi k^{4}(n-1)^{4}R^{6}ANl}{9}+\frac{i8\pi k^{5}(n-1)^{5}R^{7}ANl}{105}
	\end{aligned}
\end{equation}
Let
\begin{equation}\label{eq.23}
	a=\frac{4\pi k(n-1)R^{3}AN}{3}
\end{equation}
\begin{equation}\label{eq.24}
	b=\frac{6\pi k^{2}(n-1)^{2}R^{4}AN}{2}
\end{equation}
\begin{equation}\label{eq.25}
	c=\frac{8\pi k^{3}(n-1)^{3}R^{5}AN}{15}
\end{equation}
\begin{equation}\label{eq.26}
	d=\frac{2\pi k^{4}(n-1)^{4}R^{6}AN}{9}
\end{equation}
\begin{equation}\label{eq.27}
	e=\frac{8\pi k^{5}(n-1)^{5}R^{7}AN}{105}
\end{equation}
Substitute equations (\ref{eq.23})-(\ref{eq.27}) into (\ref{eq.22}), we have
\begin{equation}\label{eq.28}
	 u_{O_{l}}=Ae^{ikl}(ial-bl-icl+dl+iel)
\end{equation}
Adding the amplitude $Ae^{ikl}$ of the incident light at point $O_l$ and Eq. (\ref{eq.28}), the total amplitude at point $O_l$ can be obtained as 
\begin{equation}\label{eq.29}
	U_{O_{l}}=Ae^{ikl}+ u_{O_{l}}=Ae^{ikl}(1+ial-bl-icl+dl+iel)
\end{equation}
The optical medium in Fig.\ref{Fig.3} is very thin, that is, $l$ in Equation (\ref{eq.29}) is very small, then:
\begin{equation}\label{eq.30}
	1+ial-bl-icl+dl+iel=e^{ial-bl-icl+dl+iel}=e^{i(a-c+e)l-(b-d)l}
\end{equation}
Substitute (\ref{eq.30}) into (\ref{eq.29}):
\begin{equation}\label{eq.31}
		U_{O_{l}}=Ae^{ikl}e^{i(a-c+e)l-(b-d)l}=Ae^{-(b-d)l}e^{i(k+a-c+e)l}
\end{equation}
The influence of a layer of medium with a thickness of $l$ on an incident light can be expressed as the transmittance $T$:
\begin{equation}\label{eq.32}
	U_{O_{l}}=Ae^{-(b-d)l}e^{i(k+a-c+e)l}=AT
\end{equation}
where
\begin{equation}\label{eq.33}
	T=e^{-(b-d)l}e^{i(k+a-c+e)l}
\end{equation}

\section{Thick medium dispersion properties}
\begin{figure}[H] 
	\centering 
	\includegraphics[width=0.7\textwidth]{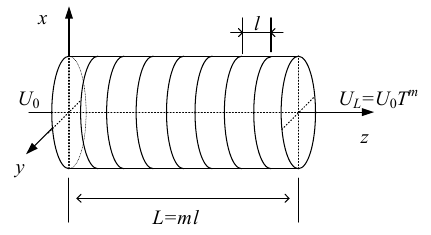} 
	\caption{Replace medium of length $L$ by $m$ layers of thin medium} 
	\label{Fig.4} 
\end{figure}

A macro medium with a length of $L$, as shown in Fig.\ref{Fig.4}, can be equivalent to $m$ layers of thin medium with a thickness of $l$. Then we have

\begin{equation}\label{eq.34}
	\begin{aligned}
		U_L&=U_{0}T^{m}=U_{O}[e^{-(b-d)l}e^{i(k+a-c+e)l}]^{m}=U_{0}e^{-(b-d)ml}e^{i(k+a-c+e)ml}	\\
		&=U_{0}e^{-(b-d)L}e^{i(k+a-c+e)L}=U_{0}e^{ik^{\prime}L}
	\end{aligned}
\end{equation}
where
\begin{equation}\label{eq.35}
	k^{\prime}=i(b-d)+(k+a-c+e)
\end{equation}
It can be seen from Eq. (\ref{eq.34}) that when light propagates in a non-uniform medium, the real part $k+a-c+e$ of the wave number $k^{\prime}$ determines the wavelength $\lambda=2\pi /(k+a-c+e)$ of the light, and the imaginary part $i(b-d)$ determines the attenuation of the light. Assume that the optical wavelength in a uniform medium is $\lambda$, and its frequency is $\nu =v_{0}/\lambda$, then the propagation velocity of light in a dispersive medium can be obtained from the real part of wave number as:
\begin{equation}\label{eq.36}
	v=\nu \lambda^{\prime}=\frac{v_{0}}{\lambda}\frac{2\pi}{k+a-c+e}=v_{0}\frac{k}{k+a-c+e}
\end{equation}
According to Equation (\ref{eq.36}), the refractive index of the medium is:
\begin{equation}\label{eq.37}
	n(\lambda)=\frac{v_0}{v}=\frac{k+a-c+e}{k}=1+\frac{a}{k}-\frac{c}{k}+\frac{e}{k}
\end{equation}
Substitute (\ref{eq.23}), (\ref{eq.25}), and (\ref{eq.27}) into (\ref{eq.37}), it can be obtained
\begin{equation}\label{eq.38}
	\begin{aligned}
	n(\lambda)&=1+\frac{4\pi (n-1)R^{3}AN}{3}-\frac{8\pi k^{2}(n-1)^{3}R^{5}AN}{15}+\frac{8\pi k^{4}(n-1)^{5}R^{7}AN}{105}\\
	&=1+\frac{4\pi(n-1)R^{3}AN}{3}-\frac{1}{\lambda^2}\frac{32\pi^{3}(n-1)^{3}R^{5}AN}{15}+\frac{1}{\lambda^4}\frac{128\pi k^{4}(n-1)^{5}R^{7}AN}{105}\\
	&=n_{0}-\frac{a^{\prime}}{\lambda^2}+\frac{b^{\prime}}{\lambda^{4}}
	\end{aligned}
\end{equation}
where
\begin{equation}\label{eq.39}
    \begin{aligned}
	&n_0=1+\frac{4\pi(n-1)R^{3}AN}{3}\\
	&a^{\prime}=\frac{32\pi^{3}(n-1)^{3}R^{5}AN}{15}\\
	&b^{\prime}	=\frac{128\pi k^{4}(n-1)^{5}R^{7}AN}{105}
	\end{aligned}
\end{equation}
It is obvious that equation (\ref{eq.38}) is the dispersion law for inhomogeneous media, which is consistent with the Cauchy dispersion equation.
\section{Conclusion}
Using Huygens wavelet as the fundamental element of light field, the interaction between light and particles can be described correctly, and the dispersion caused by medium inhomogeneity can be found. The method using Huygens' principle to study dispersion can also be extended to the atomic scale, as well as being used to study the relationship between optical materials and dispersion.

\newpage
\small
\bibliographystyle{unsrt}
\bibliography{ref}

\end{document}